\documentclass[twocolumn,pre,aps,showpacs]{revtex4}
\usepackage{bm}
\def\dt{\Delta t}
\def\dx{\Delta x}
\def\pt{\partial_t}
\def\grad{\nabla}
\def\frac#1#2{{\textstyle{#1\over #2}}}
\def\b#1{{\bf #1}}
\def\lk{\lambda_{\b k}}
\def\bphi{\bm{\phi}}
\def\d{\delta}
\input epsf

\begin{document}
\title{Fast and Accurate Coarsening Simulation with an Unconditionally 
Stable Time Step}
\author{Benjamin P. Vollmayr-Lee} 
\email{bvollmay@bucknell.edu}
\affiliation{Department of Physics, Bucknell University, Lewisburg PA
  17837, USA}
\author{Andrew D. Rutenberg}
\email{andrew.rutenberg@dal.ca}
\homepage{http://www.physics.dal.ca/~adr}
\affiliation{Department of Physics and Atmospheric Science, Dalhousie
  University, Halifax, Nova Scotia, Canada B3H 3J5}
\date{August 8, 2003}
\begin{abstract}
We present Cahn-Hilliard and Allen-Cahn numerical integration
algorithms that are unconditionally stable and so provide
significantly faster accuracy-controlled simulation.  Our stability
analysis is based on Eyre's theorem and unconditional von Neumann
stability analysis, both of which we present.  Numerical tests confirm
the accuracy of the von Neumann approach, which is straightforward and
should be widely applicable in phase-field modeling.  We show that
accuracy can be controlled with an unbounded time step $\Delta t$ that
grows with time $t$ as $\Delta t \sim t^{\alpha}$.  We develop a
classification scheme for the step exponent $\alpha$ and demonstrate
that a class of simple linear algorithms gives $\alpha=1/3$.  For this
class the speed up relative to a fixed time step grows with the linear
size of the system as $N/\log N$, and we estimate conservatively that
an $8192^2$ lattice can be integrated $300$ times faster than with the
Euler method.
\end{abstract}

\maketitle

%%%%%%%%%%%%%%%%%%%%%%%%%%%%%%%%%%%%%%%%%%%%%%%%%%%%%%%%%%%%%%%%%%%%%%%
\section{Introduction}
\label{sec:intro}

A starting point in the analysis of coarsening systems, such as the
phase separation dynamics following a quench from a disordered to an
ordered phase, is the characterization of the asymptotic {\it
late-time\/} behavior.  Most coarsening systems exhibit asymptotic
dynamical scaling with the characteristic length scale $L(t)$ given by
the size of individual ordered domains.  The growth-law $L \sim
t^n$ is determined by only a few general features, such as
conservation laws and the nature of the order parameter (see
\cite{Bray94} for a review).  For conserved Cahn-Hilliard equations
describing phase-separation, $L \sim t^{1/3}$ at late times.  More
detailed information about the scaling state is difficult to obtain
analytically.  Indeed the very existence of scaling has only been
demonstrated empirically in simulations and experiments.
Consequently, computer simulations of coarsening models, especially
phase-field type models like the Cahn-Hilliard equation, play an
essential role in our understanding and characterization of late-stage
coarsening.

These simulations face several restrictions.  To accurately resolve the
asymptotic structure it is necessary to evolve until late times so
that $L(t) \gg w$, where $w$ is the domain wall width.  However,
to avoid finite-size effects we must halt the simulation when $L(t)$
is some fraction of the system size $L_{\rm sys}$.  Additionally, to
resolve the domain wall adequately the lattice spacing $\dx$ must be
be sufficiently small compared to the domain wall width $w$.  Very
large lattices of linear size $L_{\rm sys}/\dx$ are necessary to
satisfy all of these requirements: $\dx < w \ll L(t) <
L_{\rm sys}$.  Accurate studies of the scaling state require us to
evolve large systems to late times.

Unfortunately, current computational algorithms are very inefficient
in their time integration.  The standard Euler integration of the
Cahn-Hilliard (CH) and Allen-Cahn (AC) coarsening models, for
conserved and non-conserved dynamics, respectively, is known to be
unstable for time steps $\Delta t$ above a threshold fixed by the
lattice spacing $\Delta x$ --- this is the ``checkerboard''
instability \cite{Rogers88}.  This imposes a fixed time step
irrespective of the natural time scale set by the physical dynamics.
The domain walls move increasingly slowly, for example, the CH
equation yields asymptotic domain wall velocities $v \sim \partial
L/\partial t \sim t^{-2/3}$.  Consequently, a fixed time step results in
ever-decreasing amounts of domain wall motion per step and eventually
becomes wastefully accurate.

Ideally, one would like a stable integration algorithm, which would
allow {\it accuracy} requirements rather than {\it stability}
limitations to determine the integration step size.  Recently, Eyre
proved the existence of unconditionally gradient stable algorithms
(essentially a strict non-increase in free energy for every possible
time step) \cite{Eyre98a}, and provided explicit examples of stable
steps for both CH and AC dynamics \cite{Eyre98a,Eyre98b}.  The present
work is concerned with developing these methods in two directions:
clarifying and expanding the class of unconditionally stable
algorithms, and deriving the accuracy limitations on these algorithms.

\begin{figure}[htb]
\centerline{
\epsfxsize=3.0in
\epsfbox{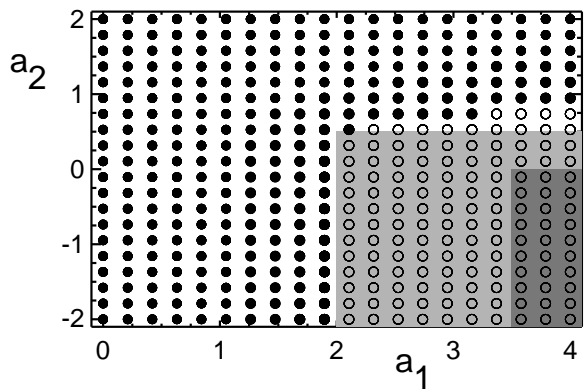}
}
\caption{For time steps parametrized as in (\protect\ref{eq:CHdirect})
the dark shaded region indicates parameters for which Eyre's theorem
proves unconditional gradient stability, while the light shaded region
corresponds to unconditional von Neumann (linearly) stable steps.  The
open circles denote steps that are numerically gradient stable under
all of our tests, as described in
Sec.~\protect\ref{subsec:gradientstable}, while the black circles
indicate parameters that were found numerically not to be gradient
stable.}
\label{fig:abstab}
\end{figure}

Our main results for {\it stability} are the following.  We have
determined the parameter range for which
Eyre's theorem proves unconditional gradient stability
(Sec.~\ref{subsec:Eyrestability}), and we present Eyre's theorem in
appendix \ref{app:Eyre}.  We have also determined the parameter range
that is unconditionally von Neumann (vN) stable, that is, linearly
stable for any size time step (Sec.~\ref{subsec:vonNeumann}).  The
latter range is a superset of the former, and neither appear to have
been previously determined.  We have also performed numerical tests of
stability in dimension $d=2$ (Sec.~\ref{subsec:gradientstable}) and
found that the vN stability condition appears to be sufficient for
identifying unconditionally gradient stable steps.  Specifically, for
the parameterless form of the CH equation (see \cite{Bray94})
\begin{equation}
        \dot\phi =  - \grad^2 ( \grad^2\phi + \phi - \phi^3), 
	\label{eq:CH} 
\end{equation}
there exists a class of semi-implicit steps
\begin{eqnarray}
        \tilde\phi_{t+\dt} &+& \dt \grad^2 [ (1-a_1)\tilde\phi_{t+\dt} 
                + (1-a_2) \grad^2 \tilde\phi_{t+\dt} ]
                \nonumber  \\ 
        = \phi_t &+& \dt \grad^2[ -a_1 \phi_t - a_2  \grad^2\phi_t + 
                 \phi_t^3].
        \label{eq:CHdirect}
\end{eqnarray}
that may be solved for the updated field $\tilde\phi_{t+\dt}$
efficiently by means of fast Fourier transform (FFT).  The various
stability conditions for these steps are depicted in terms of $a_1$
and $a_2$ in Fig.~\ref{fig:abstab}.  The stability conditions do not
depend on the lattice type or dimension, on the volume fraction, or on
the form of the lattice Laplacian.  This implies, for example, that
these algorithms could be combined with adaptive mesh techniques (see,
for example, \cite{Provatas98}) for independent control of spatial and
temporal discretization.  Fig.~\ref{fig:abstab} suggests that the
unconditional vN stability conditions, which are widely applicable and
relatively easy to analyze, may provide a reasonably accurate proxy
for unconditional gradient stability.  We have also determined the
analogous stability conditions for the AC equation.

When stability is not the limiting factor, practical limits are still
imposed by {\it accuracy}.  To maintain the domain wall profile to a
given accuracy, a time step should be chosen so that the wall only
moves a fraction of its width $w$ in a single step.  For a scaling
system with $L\sim t^n$, where $n \leq 1$ generally, the
passage time $\tau$ scales like $w/v \sim w/\dot L \sim t^{1-n}$
at late times.  Then the natural time step should scale as
\begin{equation}
        \dt_{nat} \sim \tau \sim t^{1-n}.
        \label{eq:optimal}
\end{equation} 
For CH dynamics, $n=1/3$ and $\dt_{nat}\sim t^{2/3}$ while for AC
dynamics $n=1/2$ and $\dt_{nat}\sim t^{1/2}$.  However, we show
that these stable algorithms are still not capable of accurately
simulating coarsening using the natural time scale --- despite their
stability.  For example, accuracy limits the stable CH steps given
above to ``only'' $\dt \sim t^{1/3}$.

To understand the limitations imposed on even stable algorithms by
accuracy, we study in Sec.~\ref{sec:accuracy} the truncation error for
the CH equation for general numerical algorithms, and determine the
how these terms scale with time to all orders in $\dt$
(Sec.~\ref{subeq:fieldscaling}).  We develop a classification scheme
for such algorithms based on the lowest order $p$ of $\dt^p$ at which
truncation error fails to follow its optimal scaling and show that
this term limits the accuracy of the algorithm at late times
(Sec.~\ref{subsec:pclassification}).  Our
analysis leads to the conclusion that accuracy requires a time step
\begin{equation}
\label{eq:dtp}
 \dt \sim t^{2(p-1)/3p}
\end{equation}
for the CH model.  The algorithms in Eq.~(\ref{eq:CHdirect}) have $p=2$,
meaning the error becomes sub-optimal at $O(\dt^2)$, the leading error
term.  This result is consistent with our numerical observations.  Our
simple analysis for the natural time step, Eq.~(\ref{eq:optimal}),
corresponds to the $p=\infty$ class.  We are unable to identify any 
such ``perfect'' algorithms for the CH case; they are quite
likely impossible for any nonlinear problem.

Next, we turn to the question of practical advantage.  Various
computational algorithms have been developed to mitigate the impact of
instabilities by increasing $\Delta t$ by a fixed factor compared to
the simplest Euler discretization.  For example, the
cell-dynamical-scheme (CDS) \cite{Oono88} exploits universality to
choose a free energy that is convenient in terms of numerical
stability.  More recently, Fourier spectral methods
\cite{Chen98,Zhu99} have been shown to increase the maximum $\Delta t$
by an impressive two orders of magnitude.  However, these methods
still require fixed time steps and so cannot adjust to the naturally
slowing CH dynamics.

In Sec.~\ref{sec:speedup} we determine the relative advantage of
integration by algorithms such as Eq.~(\ref{eq:CHdirect}) compared to
the conventional Euler method.  For a reasonably conservative choice
of accuracy requirements, we find for an $8192\times 8192$ lattice
(currently feasible for a linux workstation) with $\dx =1$ that the
new methods can integrate up to finite size effects roughly a factor
of $300$ times faster than possible with the Euler method.  The
advantage of unconditionally stable steps increases with larger system
sizes: for lattices of linear size $N$ we show the relative advantage
in speed is order $N/\log N$, regardless of spatial dimension of the
system.  This means that as computational power continues to increase,
unconditionally gradient stable algorithms will become even more
valuable.

We present a summary and outlook for future developments and
applications in Sec.~\ref{sec:conclusion}.

%%%%%%%%%%%%%%%%%%%%%%%%%%%%%%%%%%%%%%%%%%%%%%%%%%%%%%%%%%%%%%%%%%%%%%%%%
\section{Stability}
\label{sec:stability}

The parameterless form of the CH equation for a conserved
scalar field \cite{Bray94} is
\begin{equation}
        \dot\phi = \grad^2 \mu
	\label{eq:CH2} 
\end{equation}
where $\mu$ is the local chemical potential given by
\begin{equation}
        \mu(\b x) \equiv {\delta F \over \delta \phi(\b x)},
	\label{eq:mu}
\end{equation}
and $F[\phi]$ is the free energy functional, taken here to be
\begin{equation}
        F[\phi] \equiv \int d^dx \left[ \frac 1 2 (\grad \phi)^2 +
	\frac 1 4 (\phi^2-1)^2  \right].
	\label{eq:F}
\end{equation}
The second term in $F$ represents a double-well potential with
equilibrium values $\phi = \pm 1$, and Eqs.~(\ref{eq:CH2}),
(\ref{eq:mu}), and (\ref{eq:F}) combine to give Eq.~(\ref{eq:CH}).
The parameterless form of the AC equation \cite{Bray94}
is
\begin{equation}
	\dot\phi = -\mu = \grad^2\phi + \phi - \phi^3.
	\label{eq:AC}
\end{equation}

For dissipative dynamics such as the CH and AC equations, a discrete
time stepping algorithm is defined to be {\em gradient stable} only if
the free energy is non-increasing, $F[\tilde\phi_{t+\Delta t}] \leq
F[\phi_t]$, for any field configuration $\phi_t$.  The other
requirements for gradient stability, e.g.\ that stable fixed points
must correspond to minima of $F$, or that $F$ should increase without
bound for large $\phi$, are already manifest in the discretized forms
of these equations.  Gradient stability may reasonably be regarded as
the ultimate stability criterion for the CH equation.

{\bf Unconditional gradient stability} means that the conditions for
gradient stability hold for {\em any} size time step $\dt \in
[0,\infty)$.  Since unconditionally stable steps are our primary
concern, we will henceforth use ``stable'' or ``unstable'' to refer to
the behavior for arbitrarily large $\dt$.  That is, ``stable'' implies
unconditionally stable, while a fixed time step algorithm like the 
Euler step may be referred to as ``unstable'' or conditionally stable.

The Euler time discretization of the CH equation is
\begin{equation}
        \phi^{\rm Eu}_{t+\dt} \equiv \phi_t + \dt\, \grad^2 \mu_t.
        \label{eq:CHeuler}
\end{equation}
The Euler update is ``explicit'' since the field at the earlier time
step ($\phi_t$) explicitly determines the field at the next time step
($\phi_{t+\dt}$).  It is also unstable for values of $\dt$ that exceed
a lattice-dependent threshold, $\dt_{max} \sim \dx^4$ \cite{Rogers88}.
The fully implicit time step is obtained by replacing $\mu_t$ with
$\mu_{t+\dt}$ in Eq.~(\ref{eq:CHeuler}), and is, like the Euler step,
accurate to $O(\dt)$.  Other time steps, which involve splitting $\mu$
into parts evaluated at $t$ and at $t+\dt$, are generally called
semi-implicit methods.

Remarkably, Eyre \cite{Eyre98a,Eyre98b} proved that appropriate
semi-implicit parametrizations can lead to stable update steps for
both the CH and AC equations.  To explore these possibilities, it is
useful to introduce a general family of such steps for the CH equation
in an arbitrary spatial dimension:
\begin{eqnarray}
        \tilde\phi_{t+\dt} &+& (1-a_1) \dt \grad^2\tilde\phi_{t+\dt} 
                + (1-a_2) \dt \grad^4 \tilde\phi_{t+\dt} 
                \nonumber \\ &-& 
        (1-a_3) \dt \grad^2[\phi_t^m \tilde\phi_{t+\dt}^{3-m}] 
                = \nonumber\\
        \phi_t &-& a_1 \dt \grad^2\phi_t - a_2 \dt \grad^4\phi_t + 
                a_3 \dt \grad^2\phi_t^3.
        \label{eq:CHgeneral}
\end{eqnarray}
This reduces to Eq.\ (\ref{eq:CHdirect}) for $a_3=1$.  For each of the
three terms on the right-hand side of Eq.~(\ref{eq:CH}) there
generally are {\em both} explicit and implicit contributions to
Eq.~(\ref{eq:CHgeneral}), and this will be exploited to construct
stable dynamics for any size $\Delta t$.  For all values of the
parameters $a_i$ and $m$ this step gives a solution
$\tilde\phi_{t+\dt}$ that is order $O(\dt)$ accurate.  The implicit
terms are denoted $\tilde\phi_{t+\dt}$, with $\phi_{t+\dt}$ reserved
to represent the exact field obtained by integration of
Eq.~(\ref{eq:CH}) over the time step $\dt$.  We choose our
parameterization such that $a_1=a_2=a_3=1$ corresponds to the Euler
update Eq.~(\ref{eq:CHeuler}), while $a_1=a_2=a_3=0$ is the fully
implicit step.  For $a_3 \neq 1$ we have, motivated by Eyre, a mixed
non-linear term with $0 \leq m<3$ that combines implicit and explicit
terms.

It is useful to sort algorithms described by Eq.~(\ref{eq:CHgeneral})
into three categories based on how they are implemented numerically.
First, when $a_3=1$ we have {\bf linear direct} steps, where the
equation for $\tilde\phi_{t+\dt}$ is linear and has spatially uniform
coefficients so the updated field can be found efficiently with FFT
methods.  Second, when $a_3 \neq 1$ but $m=2$ then the implicit
equation remains linear in $\tilde\phi_{t+\dt}$ but no longer has
spatially uniform coefficients.  Eyre outlines an iterative procedure
for solving these equations \cite{Eyre98b}, so we call these {\bf
linear iterative} steps.  Insisting on convergence of the iterative
procedure restricts this class to a subset of parameter values.
Finally, for $a_3\neq 1$ and $m\neq 2$ the update equation is {\bf
nonlinear}.  For some parameter values the nonlinear equation can
lead, unphysically, to multiple solutions.  This occurs for both the
fully implicit case $a_1=a_2=a_3=m=0$, as well as the Crank-Nicholson
case $a_1=a_2=a_3=1/2$, $m=0$, whenever $\dt$ exceeds a threshold
value \cite{Eyre98a}.  Generally the nonlinear equations require
solution by the Newton-Raphson method, which is complicated to
implement in two or more spatial dimensions.  For some parameter
values this can be demonstrated to be absolutely convergent, so
nonlinear steps provide a viable option --- though not one we have
explored numerically.

The step parametrization for the AC equation analogous to
Eq.~(\ref{eq:CHgeneral}) is
\begin{eqnarray}
        \tilde\phi_{t+\dt} &-& (1-a_1) \dt\tilde\phi_{t+\dt} 
                - (1-a_2) \dt \grad^2 \tilde\phi_{t+\dt} 
                \nonumber \\ &+& 
        (1-a_3) \dt [\phi_t^m \tilde\phi_{t+\dt}^{3-m}] 
                = \nonumber\\
        \phi_t &+& a_1 \dt \phi_t + a_2 \dt \grad^2\phi_t -
                a_3 \dt \phi_t^3,
        \label{eq:ACgeneral}
\end{eqnarray}
which we include because the theoretical stability analysis follows
nearly identically for the CH and AC equations, and the stability
regions are given by the same shaded regions of Fig.~1.

%%%%%%%%%%%%%%%%%%%%%%%
\subsection{Unconditionally Stable Steps from Eyre's Theorem} 
\label{subsec:Eyrestability}

Eyre's theorem (see appendix \ref{app:Eyre}) shows that an
unconditionally gradient
stable algorithm results, for both the CH and AC equations, if one can
split the free energy appropriately into {\it contractive} and {\it
expansive} parts, $F=F^C + F^E$, and treat the contractive parts
implicitly and the expansive parts explicitly.  That is, the
CH equation (\ref{eq:CH2}) is discretized as
\begin{equation}
  \tilde\phi_{t+\dt} - \dt\grad^2\mu^C_{t+\dt} = \phi_t + \dt\grad^2\mu^E_t,
	\label{eq:Eyrestep}
\end{equation}  
while the AC equation (\ref{eq:AC}) is discretized as
\begin{equation}
  \tilde\phi_{t+\dt} + \dt\mu^C_{t+\dt} = \phi_t - \dt\mu^E_t,
	\label{eq:Eyrestep2}
\end{equation}  
where $\mu_i^X = \partial F^X/\partial\phi_i$ for lattice site $i$,
and where $\grad^2$ implies a lattice laplacian.  The necessary
condition on the splitting is the same for both equations and 
may be stated by introducing the Hessian matrices
\begin{equation}
M_{ij} = {\partial^2 F\over\partial\phi_i\partial\phi_j}, \qquad
M^E_{ij} = {\partial^2 F^E\over\partial\phi_i\partial\phi_j}, \qquad
M^C_{ij} = {\partial^2 F^C\over\partial\phi_i\partial\phi_j},
\end{equation}
where $i,j$ denote lattice sites.  First, we must have all eigenvalues
of $\b M^E$ non-positive and all eigenvalues of $\b M^C$ non-negative.
Second, as shown in appendix \ref{app:Eyre}, for $\lambda_{min}$ equal
to the smallest eigenvalue of $\b M$ and $\lambda^E_{max}$ the largest
eigenvalue of $\b M^E$, we need
\begin{equation}
  \lambda^E_{max} \leq \frac 1 2 \lambda_{min}.
\label{eq:Eyrecondition}
\end{equation}
This also automatically satisfies the convexity requirement for $\b
M^E$, since $\lambda_{min} < 0$.

To identify the appropriate splittings, it is useful to break the free
energy Eq.~(\ref{eq:F}), in its lattice-discretized form, into three
parts (neglecting the irrelevant constant $V/4$ term):
\begin{eqnarray}
F^{(1)} &=& -\sum_i\,\frac 1 2\phi_i^2, \qquad
F^{(2)} = \sum_i \,\frac 1 2 (\nabla\phi)_i^2, \qquad \nonumber\\
F^{(3)} &=& \sum_i\,\frac 1 4\phi_i^4 \qquad.
\end{eqnarray}
with corresponding Hessian matrices $\b M^{(i)}$.  The first,
$M^{(1)}_{ij} = -\delta_{ij}$, where $\delta_{ij}$ is the Kronecker
$\delta$-function, has all eigenvalues equal to $-1$.  Next,
$M^{(2)}_{ij} = (-\grad^2)_{ij}$ is negative the lattice laplacian,
which can always be diagonalized by going to Fourier space.  It
immediately follows that the eigenvalues of $\b M^{(2)}$ are strictly
non-negative.  (Even for irregular spatial discretizations, the $\b
M^{(2)}$ eigenvalues must be non-negative.)  Finally, $M^{(3)}_{ij} =
3\phi_i^2 \delta_{ij}$, which has strictly non-negative eigenvalues as
well.  We parametrize the splitting via
\begin{equation}
	F^E = \sum_{i=1}^3 a_i F^{(i)} \qquad 
	F^C = \sum_{i=1}^3 (1-a_i) F^{(i)}
	\label{eq:Eyresplitting}
\end{equation}
which results in the general CH step Eq.~(\ref{eq:CHgeneral}) and AC step
Eq.~(\ref{eq:ACgeneral}) when $m=0$.

Now to obtain bounds: since the sum of matrices, $\b M = \b M^{(1)}+
\b M^{(2)}+ \b M^{(3)}$, has eigenvalues bounded by the sum of the
bounds, the minimum eigenvalue of $\b M$ satisfies
$\lambda_{min} \geq -1$.  Therefore Eq.\ (\ref{eq:Eyrecondition}) is
satisfied by ensuring $\lambda^E_{max} \leq -1/2$.

One example that satisfies these conditions is the splitting $F^E =
F^{(1)}$ and $F^C = F^{(2)} + F^{(3)}$, since $\lambda^E_{max}=-1$
satisfies Eq.~(\ref{eq:Eyrecondition}) and $\b M^C$ has strictly
non-negative eigenvalues.  This provides a gradient stable {\bf
nonlinear} step with $a_1=1$ and $a_2 = a_3 = 0$.  This case was
identified by Eyre \cite{Eyre98a}, who noted that the convexity
requirement for the splitting guarantees absolute convergence of the
Newton-Raphson method.

Eyre also presents a technique for identifying stable linear {\bf
direct} algorithms \cite{Eyre98a}, which relies on the fact that
$\phi^2$ is bounded.  It exceeds unity only slightly in the CH
equation and only in the interior region of a curved interface due to
Gibbs-Thompson effects \cite{subsaturation}.  Therefore the
eigenvalues of $\b M^{(3)}$ have an effective upper bound,
approximately three.  If we then take $F^E = a_1 F^{(1)} + F^{(3)}$
(so $a_3=1$ and $a_2=0$) the eigenvalues of $\b M^E$ are of the form
$-a_1 + 3\phi_i^2$ and satisfy Eq.~(\ref{eq:Eyrecondition}) for
$\phi_i^2\lesssim 1$ if $a_1\gtrsim 7/2$.  Any value $a_2\leq 0$ will
give the same result, since negative values of $a_2$ can only decrease
the eigenvalues of $\b M^E$.  These choices imply $F^C =
(1-a_1)F^{(1)} + (1-a_2)F^{(2)}$, which has the necessary non-negative
eigenvalues for the range of $a_1$ and $a_2$ given above.  Therefore
we can identify a class of gradient stable {\bf direct} CH and AC
steps as
\begin{equation}
	 a_1 \gtrsim 7/2 \qquad a_2 \leq 0 \qquad a_3 = 1.
	\label{eq:Eyrestability}
\end{equation}
This gives the dark gray shaded region in Fig.~\ref{fig:abstab}.
These represent {\it sufficient} restrictions on the $a_i$ to satisfy
the conditions for Eyre's theorem; however other values of the $a_i$
may be gradient stable as well.  

Eyre provided specific step examples for all three implementation
categories: a {\it nonlinear} step, with $a_1=1$, $a_2=a_3=0$, and
$m=0$, a linear {\it iterative} step with $m=2$ and the same $a_i$ as
the nonlinear step, and a linear {\it direct} step with $a_1=3$,
$a_2=0$, and $a_3=1$ \cite{Eyre98a,Eyre98b}.  The nonlinear step is the
example presented earlier in this section, and its gradient stability
follows from Eyre's theorem.  However, it is not clear to us that
Eyre's theorem can be directly applied to the iterative steps, and in
fact we find Eyre's proposed iterative method to be numerically
unstable, as described in Sec.~\ref{subsec:gradientstable}.
Finally, the $a_1$ value in the direct step violates
Eq.~(\ref{eq:Eyrestability}), so this case does not follow from Eyre's
theorem.

%%%%%%%%%%%%%%%%%%%%%%
\subsection{Unconditional von Neumann Stability}
\label{subsec:vonNeumann}

Since Eyre's theorem provides, in principle, only a subset of the
possible gradient stable steps, complementary approaches for
determining stability are desirable.  In this section we extend von
Neumann's (vN) linear stability analysis \cite{NumericalRecipes} to
arbitrary time steps, which we call unconditional vN stability.  Since
any gradient stable algorithm is likely also linearly stable, the
von Neumann analysis would appear to identify a {\em superset} of
possibly gradient stable algorithms: in principle the vN analysis
could also identify some unwanted {\em non-linearly} unstable
algorithms.  As shown in Fig.~\ref{fig:abstab}, though, the 
vN stability boundary corresponds quite well with the numerically
determined gradient stability line.  This leads us to suggest that the
approach of imposing unconditional vN stability on a broadly parametrized
class of semi-implicit algorithms, followed by numerical checking,
could be fruitfully adapted to a wide variety of applications.

We analyze the general step Eq.~(\ref{eq:CHgeneral}) for linear
stability around a constant phase $\phi=\phi_0$.  It is important to
realize there are {\em physical}, and therefore desirable, linear
instabilities in the continuum CH and AC equations.  Therefore it is
important to distinguish between these and the unphysical
instabilities induced by the numerical implementation.  Take 
$\phi(\b x,t) \equiv \phi_0 + \eta(\b x, t)$, and linearize the
CH equation (\ref{eq:CH})
in $\eta$ to get $\dot\eta = -\grad^2 (\grad^2\eta + \eta - 3 \phi_0^2
\eta)$.  Fourier transform this to get
\begin{eqnarray}
        \dot\eta_{\b k} &=& 
		-\lk  ( \lk+ k_0^2 ) \eta_{\b k}, \label{eq:CHlin} \\
                k_0^2 &\equiv& 1-3\phi_0^2.
\end{eqnarray}
Here $\lk$ is the eigenvalue of the Laplacian and is non-positive,
with $\lambda_{kmin} \leq \lk \leq 0$ (note that $\lk = -k^2$ in the
continuum).  The minimum value $\lambda_{kmin}$ depends on the
lattice, spatial dimension, and specific form of the laplacian.
Similarly, for the same $\phi$ linearize the AC equation (\ref{eq:AC})
in $\eta$ and Fourier transform to get
\begin{equation}
   	\dot\eta_{\b k} =  (\lk+ k_0^2 ) \eta_{\b k}.
	\label{eq:AClin} 
\end{equation}
The physical instability for both Eqs.~(\ref{eq:CHlin}) and (\ref{eq:AClin})
occurs for
\begin{equation}
	-\lk < k_0^2, 
	\label{eq:spinodal}
\end{equation}
which corresponds in the CH equation to spinodal decomposition
\cite{Bray94}.  We stress that while these Fourier modes are linearly
unstable, the dynamics of spinodal decomposition is {\em gradient
stable} and represents a physical decrease of the free energy, which
is why it must be retained.

We now linearize and Fourier transform our general CH step
Eq.~(\ref{eq:CHgeneral}) as above to get
\begin{eqnarray}
        [1-\lk \dt \{ && (a_1-1)-\lk (1-a_2) \nonumber \\ 
                        && +\phi_0^2 (1-a_3) (3-m)\}] \eta_{\b k,t+\dt} = 
	\nonumber \\
\ \  %%% needed due to LaTex bug
        [1-\lk \dt \{ && a_1 + \lk a_2  \nonumber \\
                          && + \phi_0^2 (-3 a_3 +m(a_3-1)) \}] \eta_{\b k,t}
        \label{eq:lin_general}
\end{eqnarray}
Writing this as 
\begin{equation}
	[1+\dt{\cal L}_{\bf k}]\eta_{\b k,t+\dt} = 
	[1+\dt{\cal R}_{\bf k}]\eta_{\b k,t},
	\label{eq:vNeq}
\end{equation}
the von Neumann stability criterion is 
\begin{equation} 
        | 1 + \dt{\cal L}_{\bf k} | > | 1 + \dt{\cal R}_{\bf k} |, 
        \label{eq:vNdef}
\end{equation}
which implies that small deviations from the constant solution evolve to
decrease in magnitude.  We want to impose this stability condition for
all ${\bf k}$ and {\em arbitrary} positive $\dt$.  For large $\dt$,
Eq.~(\ref{eq:vNdef}) implies $|{\cal L}_{\bf k}| > |{\cal R}_{\bf k}|$.
The left-hand side of Eq.~(\ref{eq:vNdef}) can be made to
violate the inequality for small $\dt$ unless ${\cal L}_{\bf k} \geq 0$. 
Combining these conditions we have
\begin{equation}
        {\cal L}_{\bf k} > | {\cal R}_{\bf k} |,
        \label{eq:vNunconditional}
\end{equation}
which is a {\it necessary and sufficient} condition for unconditional
linear stability.  This condition applies to {\em all} first-order
time steps that can be expressed in the form given by
Eq.~(\ref{eq:vNeq}).

We examine the linear stability condition in two steps.  First,
${\cal L}_{\bf k}>{\cal R}_{\bf k}$:
\begin{equation}
        0< {\cal L}_{\bf k} - {\cal R}_{\bf k}  = (-\lk)[-1-\lk+3\phi_0^2]. 
	\label{eq:vNspinodal}
\end{equation}
This reduces to the spinodal condition, Eq.~(\ref{eq:spinodal}).  Note
that all the parameters $(a_1,a_2,a_3,m)$ are absent from
Eq.~(\ref{eq:vNspinodal}), so we cannot interfere with the spinodal
condition.  This evidently follows from having a first-order accurate
step.  Next, we check for ${\cal L}_{\bf k}> -{\cal R}_{\bf k}$, which
gives
\begin{equation}
        2a_1-1 -[(3-m)(2a_3-1)+m]\phi_0^2+\lk (2a_2-1) > 0.
        \label{eq:vNcondition}
\end{equation}
If we choose $a_2 < 1/2$, then since $\lk\leq 0$ we get $2a_1-1
-[(3-m)(2a_3-1)+m]\phi_0^2 > 0$.  For $a_2 > 1/2$ we obtain a
lattice-dependent condition, that is, our inequality would contain
$\lambda_{kmin}$.

We choose to restrict ourselves to lattice-{\em independent} stability
conditions as these are more practical: they carry over into any
lattice or spatial dimension.  For this purpose we take $a_2<1/2$.
This gives the vN stable conditions
\begin{eqnarray}
        a_2 &<& 1/2 \nonumber \\
        a_1 &>& {1 +\max[0,(3-m)(2 a_3-1)+m]\over2}.
\label{eq:vNstability}
\end{eqnarray}
We have let $\phi_0^2$ vary in the late-time asymptotic range of
$\phi_0^2 \in [0,1]$, where Gibbs-Thompson induced supersaturation has
be ignored, and have imposed on $a_1$ the most restrictive value that
results.  For this reason algorithms near the stability boundaries
should be avoided at early times.

For direct steps, with $a_3=1$, the second condition in
Eq.~(\ref{eq:vNstability}) becomes $a_1 > 2$.  This gives the lightly
shaded region in Fig.~\ref{fig:abstab}.  The Euler update, with
$a_1=a_2=a_3=1$ is clearly unstable since $a_2>1/2$ {\em and} $a_1<2$.
For linear iterative steps, with $m=2$, Eq.~(\ref{eq:vNstability})
becomes $a_1 > \max[1/2,a_3+1]$.  The stability condition of the
general nonlinear step cannot be further simplified from
Eq.~(\ref{eq:vNstability}), but the special case $m=0$ gives $a_1 >
\max[1/2,3a_3-1]$.

There is another special case for which the stability conditions can be
imposed, namely when $m=0$ and $a_1=a_2=a_3\equiv a$.  In this case
the vN condition Eq.~(\ref{eq:vNcondition}) becomes
\begin{equation}
        (1-2a)[-1 + 3\phi_0^2 +\lk] > 0.
	\label{eq:equala}
\end{equation}
The square brackets term is again the spinodal condition and should be
positive for all physically stable modes, so for $a < 1/2$ both vN
stability conditions reduce to the spinodal condition.  However, these
steps, which include the marginal Crank-Nicholson case ($a=1/2$) and
the stable fully implicit step ($a=0$) suffer from having multiple
solutions to the nonlinear implicit equation whenever $\dt$ exceeds
some threshold, making them unsuitable.

Regarding Eyre's proposed steps, introduced at the end of
Sec.~\ref{subsec:Eyrestability}, we note that the direct step is
 vN stable, the iterative step is marginal for vN
stability, and the nonlinear step, which was gradient stable by Eyre's
theorem, is also vN stable.

The same linearization for the general AC step Eq.~(\ref{eq:ACgeneral})
results in the same linearized equation (\ref{eq:lin_general}) but
with the substitution $-\lk \dt\to \dt$.  Since $-\lk \geq 0$, the
vN stability analysis for the AC equation is identical and also results in
Eq.~(\ref{eq:vNstability}).

%%%%%%%%%%%%%%%%%%%%%%%%%%%%%%%%%
\subsection{Numerical Stability Tests}
\label{subsec:gradientstable}

The vN stability analysis yields a considerably larger parameter range
for stable steps, Eq.~(\ref{eq:vNstability}), than those which are
provably stable by Eyre's theorem, e.g.\ Eq.~(\ref{eq:Eyrestability}).
Here we determine numerically which step parametrizations are gradient
stable, for purposes of comparison with the theoretical results.  We
focus primarily on {\bf direct} steps, with $a_3=1$, since these are
an important practical class of steps.  We consider only symmetric
quenches of the CH equation in this section, with
$\langle\phi\rangle=0$.

The primary result, shown in Fig.~\ref{fig:abstab}, is obtained as
follows.  We evolved a uniformly distributed $20 \times 20$ array of
direct CH steps with the parameter values $a_1\in(0,4)$ and
$a_2\in(-2,2)$ on a $512^2$ lattice to a final time $t_{max}$.  We
take lattice spacing $\dx=1$ here and throughout.  At regular
intervals during the evolution we tested a single direct step with $0
< \dt <10^{10}$.  This step was only used for stability testing, and
did not contribute to the time evolution.  Steps larger than $\dt =
10^{10}$ were not employed, to avoid spurious roundoff error
effects. Any system that {\em ever} increased its free energy was
labeled unstable, and plotted in Fig.~\ref{fig:abstab} with a filled
circle.  The systems were evolved in time with multiple methods.
First, we used Euler updates ($\dt =0.05$) evolved to $t_{max}=10^4$.
Next, we evolved systems with direct updates both with fixed $\dt =
100$ and with an increasing time step $\dt = 0.05\, t^{1/3}$ (both to
$t_{max}=10^6$).

As Fig.~\ref{fig:abstab} shows, all vN stable algorithms were found
numerically to be gradient stable, and the lightly shaded region
corresponds extremely well to the gradient stable systems.  Indeed,
the vN stability boundary for $a_1$ appears to be followed quite
sharply in the numerical tests.  We do find numerical gradient
stability for a region where $a_2>1/2$: this is most likely due,
ironically, to a lattice-induced stabilization.  That is, since the
lattice laplacian $\lk$ has an implementation-dependent minimum value,
the inequality (\ref{eq:vNcondition}) may be satisfied for some
$a_2>1/2$.  Therefore we expect the precise location of this boundary
to shift slightly depending on the lattice, the spatial dimension, and
the choice of lattice laplacian, but not to cross $a_2=1/2$.

With the numerical tests described above we have tested the linear
iterative step proposed by Eyre \cite{Eyre98b} and found it to be
unstable.

\begin{figure}[htb]
\centerline{
\epsfxsize=3.0in
\epsfbox{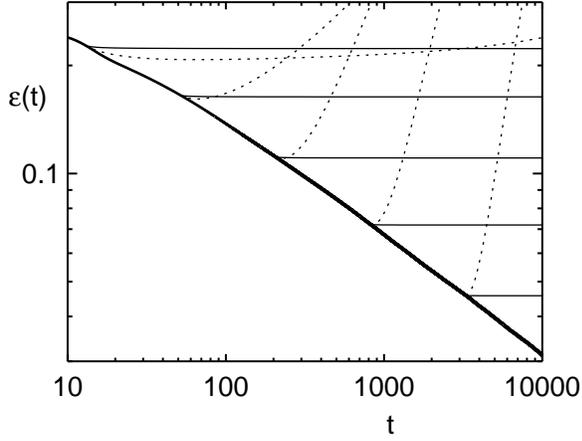}
}
\caption{Plot of the free energy density $\epsilon$ versus time (thick
solid line) approaching the asymptotic $\epsilon \sim t^{-1/3}$ decay,
as evolved with a Euler update with $\Delta t=0.01$ in a $1024^2$
system. At five distinct departure times $t_d$, separated by factors
of $4$, we show the free energies that result from a {\em single} time
step $\Delta t \in (0,10000)$, plotted versus $t=t_d+\dt$.  The dotted
lines correspond to using a common semi-implicit algorithm ($a_1=1$,
$a_2=0$, $a_3=1$) for the single step, while the thin solid lines
correspond to single steps with a vN stable direct algorithm ($a_1=3$,
$a_2=0$, and $a_3=1$).}
\label{fig:energy}
\end{figure}

\begin{figure}[htb]
\centerline{
\epsfxsize=3.0in
\epsfbox{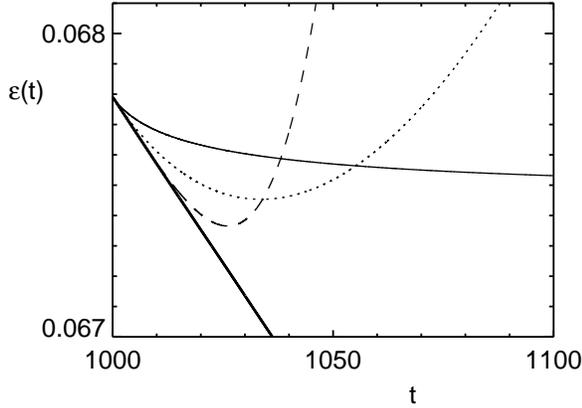}
}
\caption{As per Fig.~\protect\ref{fig:energy}, but with
$t_d=1000$. The dashed line corresponds to a single step of the Euler
update, which is gradient unstable. Both the Euler step and the
unstable semi-implicit step (dotted) are unstable under repeated steps
for much smaller $\dt$ than appear to be accurate for a single step.
}
\label{fig:enerror}
\end{figure}

To help illustrate numerical testing of gradient stability, we show a
mixture of stable and unstable steps in Figs.~\ref{fig:energy} and
\ref{fig:enerror}.  The difference between gradient stable and
unstable steps is striking: while neither are particularly accurate
for extremely large $\dt$, the unstable steps show a marked increase
in the free energy density, while the gradient stable steps adhere to
the strict non-increasing free energy condition.  However, the closer
view in Fig.~\ref{fig:enerror} shows that some cost is paid in
accuracy: for small  values of $\dt$, both the Euler step and the
unstable semi-implicit step track the physical behavior better than
the stable step.  While it may appear from Fig.~\ref{fig:enerror} that
moderately large steps may be used with unstable algorithms, this is
not case: for example using a $\dt \gtrsim 0.05$ for the Euler update
will lead to instability via accumulated error from {\it repeated}
steps.

%%%%%%%%%%%%%%%%%%%%%%%%%%%%%%%%%%%%%%%%%%%%%%%%%%%%%%%%%%%%%%%%%%%%%%%%%
\section{Accuracy}
\label{sec:accuracy}

With a gradient stable algorithm, it is possible to use a
progressively larger time step as the characteristic dynamics become
slower.  The limiting factor for the increase of the time step is then
an {\it accuracy} requirement.  

A specified accuracy criterion may be imposed on the stable steps
identified in Sec.~\ref{sec:stability} without any further theoretical
development using standard numerical adaptive step-size techniques (as
described in \cite{NumericalRecipes} and discussed in
Sec.~\ref{subsec:directerror}).  Naively, one would expect a time step
growing as $\dt \sim t^{2/3}$, for the reasons presented in
Sec.~\ref{sec:intro}.  However, this is not the case: empirically we
find significantly slower growth.  This motivated us to study the
sources of error terms in the gradient stable steps.  Our main result
is the $p$ classification scheme, which determines the allowed growth
rate of the time step according to Eq.~(\ref{eq:dtp}).

%%%%%%%%%%%%%%%%%%%%%%
\subsection{The $p$ Classification Scheme}

\label{subsec:pclassification}

We begin with an analysis of the error magnitude associated with the
various gradient stable algorithms.  The exact $\phi_{t+\dt}$,
obtained by integration of Eq.~(\ref{eq:CH}) from a given $\phi_t$,
can be expressed in terms of the fields at time $t$ by means of a
Taylor expansion:
\begin{equation}
        \phi_{t+\dt} =
	  \phi_t + \dt\, \pt\phi_t + \frac 1 2 \dt^2\, \pt^2\phi_t
          + \frac 1 {3!} \dt^3\, \partial_t^3\phi_t + \dots 
        \label{eq:CHalldt}
\end{equation}
The Euler update, Eq.~(\ref{eq:CHeuler}), is simply the truncation of
this expansion at $O(\dt)$ with resulting error $\Delta \phi^{\rm Eu}
\equiv \phi^{\rm Eu}_{t+\dt} - \phi_{t+\dt}$ given by
\begin{equation}
        \Delta \phi^{\rm Eu} =
		-\sum_{n=2}^\infty { \dt^n\over n!}
        	 \partial_t^n \phi_t.
        \label{eq:euler_error}
\end{equation}
Other step parametrizations will have different coefficients for the
$O(\dt^n)$ component of the error, but the general feature of an
expansion to all powers of $\dt$ will be the same.  Since our goal is
to have a growing time step with controlled error, successively higher
powers of $\dt$ will require coefficients decaying increasingly faster
in time.  In order to determine the limitation on how fast the time
step may grow, it is essential to know the decay rates of the
coefficients of $\dt^n$ to all orders $n$.  In this section we
demonstrate how this can be done.  We make use of the following
results for asymptotic decay rates, derived in
Sec.~\ref{subeq:fieldscaling}.  In the interfacial region (defined in
Sec.~\ref{subeq:fieldscaling})
\begin{equation}
  \partial_t^n \phi \sim t^{-2n/3} \qquad
  \partial_t^n (\grad^2)^k\phi^j \sim t^{-2n/3} 
  \label{eq:interfacedecay}
\end{equation}
whereas in the bulk, that is, all of the system not near an interface,
we find
\begin{equation}
  \partial_t^n \phi \sim t^{-(1/3)-(2/3)n} \quad
  \partial_t^n (\grad^2)^k\phi^j \sim t^{-(1/3)-2(n+k)/3} 
  \label{eq:bulkdecay}
\end{equation}

Consider first the Euler step: all the $O(\dt^n)$ coefficients are
simply proportional to the time derivative $\partial_t^n\phi$
evaluated at $t$.  If numerical stability were not a problem and we
simply increased the time step according to the naive $\dt\sim
t^{2/3}$, we would find in the interfacial region that every order in
the Taylor expansion provides an $O(t^0)$ contribution to the error,
whereas in the bulk region every order provides an $O(t^{-1/3})$
contribution.  This would present an accurate solution with a $\dt
\sim t^{2/3}$ time step, except that of course the Euler step is not
gradient stable for large time steps.

Now consider the general step, Eq.~(\ref{eq:CHgeneral}).  The error term
in this step, $\Delta\tilde\phi \equiv \tilde\phi_{t+\dt} -
\phi_{t+\dt}$ can be written as
\begin{eqnarray}
        \Delta \tilde\phi &=& 
        \Delta \phi^{\rm Eu}
        \nonumber \\ && 
        -(1-a_1) \dt \grad^2 (\tilde\phi_{t+\dt} - \phi_t ) 
        \nonumber \\ &&
        -(1-a_2) \dt \grad^4(\tilde\phi_{t+\dt} - \phi_t) 
        \nonumber \\ && 
        +(1-a_3) \dt \grad^2 
		[\phi_t^m ( \tilde\phi_{t+\dt}^{3-m} - \phi_t^{3-m})]
        \label{eq:gen_error}
\end{eqnarray}
This peculiar form with implicit $\tilde\phi_{t+\dt}$ on the right is
useful for the error analysis.  By using Eq.~(\ref{eq:CHgeneral})
iteratively, the implicit terms can be replaced by terms higher order
in $\dt$ involving the field $\phi_t$.  For example, we can derive the
$O(\dt^2)$ part of the error, using $\tilde\phi_{t+\dt}-\phi_t = \dt\,
\dot\phi_t + O(\dt^2)$ and $\tilde\phi_{t+\dt}^{3-m}-\phi_t^{3-m} =
(3-m)\dt\, \phi_t^{2-m}\dot\phi_t + O(\dt^2)$.  We find the error in
our general step to be
\begin{eqnarray}
        \Delta \tilde\phi =
        &&\Bigl[-\frac 1 2\ddot\phi_t +(a_1-1)\grad^2\dot\phi_t +
        (a_2-1) \grad^4\dot\phi_t \Bigr. \nonumber \\
        \Bigl. 
	&&+(1-a_3) (3-m) \grad^2 \phi^2_t \dot\phi_t \Bigr]\dt^2 + O(\dt^3),
        \label{eq:dt2_error}
\end{eqnarray}
where the first term comes from Eq.~(\ref{eq:euler_error}).  Now
compare the time decay of the various terms. At the interface, the
$\ddot\phi_t$ part decays as $t^{-4/3}$, but the other terms all decay
as $t^{-2/3}$.  Therefore, for general values of the $a_i$ and $m$, to
keep the $O(\dt^2)$ interfacial error fixed the time step is limited
to grow as $\dt \sim t^{1/3}$.  We see that the Euler case was special
because it made all but the first term in the $O(\dt^2)$ error vanish.
Since every term in Eq.~(\ref{eq:dt2_error}) decays faster in the bulk
than at the interface, we conclude the error is {\it interface
limited}, i.e.,\ the accuracy criterion at the interface will
determine how fast the time step can grow.  This is a generic
feature, as we will show below.

There are other ways besides using the Euler step to make the
$O(\dt^2)$ interfacial error decay as $t^{-4/3}$.  If the coefficients
satisfy
\begin{equation}
        a_1 = a_2 = 1-b \qquad  a_3 = 1-3b/(3-m).
        \label{eq:CHp3}
\end{equation}
for some $b$, then the various $\dot\phi_t$ terms
in Eq.\ (\ref{eq:dt2_error}) add to give $b\ddot\phi_t$. In this case,
\begin{equation}
  \Delta \tilde\phi = -\dt^2 \Bigl(\frac 1 2-b\Bigr)\ddot\phi_t \,\dt^2
 + O(\dt^3)
\end{equation}
and so the $O(\dt^2)$ coefficient decays as $t^{-4/3}$ at the
interface, and faster in the bulk.  From this example we can construct
the $p$ classification scheme.

Consider the truncation error term of order $\dt^n$.  This can be
obtained by iterating Eq.~(\ref{eq:gen_error}) and can be expressed as
a sum of terms of the form $\partial_t^{n-1}(\grad^2)^k\phi^j$.  If
these terms appear in the right proportions, they combined via
Eq.~(\ref{eq:CH}) to become proportional to $\partial_t^n\phi$, which
decays faster by a factor of $1/t^{2/3}$ at the interface.  This is
exactly what occurs in the $n=2$ case above when Eq.~(\ref{eq:CHp3})
is satisfied.

Now consider some value $p\geq 2$ for which all $\dt^n$ error terms
with $n<p$ are proportional to $\dt^n\partial_t^n\phi_t$, but at order
$m \geq p$ this breaks down into a sum of terms of the type
$\dt^m\partial_t^{m-1} (\grad^2)^k\phi_t^j$.  In this case the order
$p$ term provides the leading asymptotic error.  Focusing on
interfacial region, the order $p$ term goes as $\dt^p t^{-2(p-1)/3}$
according to the second term in Eq.~(\ref{eq:interfacedecay}).
Choosing the time step to hold this term at constant error would
require $\dt\sim t^\alpha$ with $\alpha = 2(p-1)/(3p)$, as displayed in
Eq.~(\ref{eq:dtp}).  Now we show that all higher- and lower-order terms
in $\dt$ will decay faster than the $\dt^p$ term for this choice of
$\alpha$.  For $n<p$, we have from the first term in
Eq.~(\ref{eq:interfacedecay}) $\dt^n t^{-2n/3}\sim t^{n(\alpha -
2/3)}= t^{-2n/3p}$, so the $n<p$ terms give ever-decreasing
contributions to the error.  For $m>p$ the error terms are of the form
$\dt^m t^{-2(m-1)/3}\sim t^{-2(m-p)/(3p)}$ which decay as well.  Hence
the asymptotic interfacial error is given by the $O(\dt^p)$ term as
advertised, and is order $t^0$.  Note that for this interface limited
$\dt\sim t^\alpha$ all bulk terms to all orders have decaying error
terms, thus establishing interface limited error as a generic feature.

%%%%%%%%%%%%%%%%%%%%%%%%%%%%%
\subsection{Quantifying Error for Direct Steps}
\label{subsec:directerror}

Direct steps, with $a_3=1$ by definition and $a_1>2$, $a_2<1/2$ for
stability, fail to satisfy Eq.~(\ref{eq:CHp3}) and so all direct steps
give $p=2$ algorithms with $\dt\sim t^{1/3}$.  This means that the
asymptotic error magnitude should be given exactly by
\begin{equation}
	|\Delta\tilde\phi| = \dt^2 |(a_1-1)\grad^2\dot\phi + 
	(a_2-1)\grad^4\dot\phi|
	\label{eq:directerror}
\end{equation}
with $\dt = A t^{1/3}$. This gives a fixed amount of error at the
interface, and all higher orders of $\dt$ give decaying contributions.
Therefore, the error magnitude is proportional to $A^2$, and we can use 
numerical measurements of Eq.~(\ref{eq:directerror}) to develop
the constant of proportionality.

We determine error numerically in the usual way
\cite{NumericalRecipes}: compare the field $\phi^{(1)}$ obtained from
a single step of size $\dt$ to the field $\phi^{(2)}$ obtained from
two steps of size $\dt/2$.  It is straightforward to show that if the
true error of the step is $E \dt^2 + O(\dt^3)$, then $\phi^{(1)} -
\phi^{(2)} = (E/2) \dt^2 + O(\dt^3)$.  Since we expect exactly $\dt^2$
error, we simply take $2(\phi^{(1)}-\phi^{(2)})$ to be the true error.

In the bulk, the error decays as $t^{-2/3}$.  The interfacial error is
not decaying, but the amount of interface decays as $t^{-1/3}$, which
means the error magnitude Eq.~(\ref{eq:directerror}) averaged over the
entire system will also decay as $t^{-1/3}$, all from the interfacial
contribution.  To determine the error per lattice site in the
interfacial region, it is necessary to divide the averaged error by
the fraction of the system in the interfacial region.  We do that as
follows.  The asymptotic free energy density is given by the product
of the surface tension $\sigma$ and interface density: $\epsilon(t) =
\sigma A_{int}(t)/L_{sys}^d\sim t^{-1/3}$, where the interfacial
``area'' $A_{int}$ is a $d-1$ dimensional hypersurface, and $L_{sys}$
is the system size.  For interface width $w$,
$A_{int}(t)w/L_{sys}^d=w\epsilon/\sigma$ represents the fraction of
the system in the interfacial region.  Multiplying the averaged error
by $\sigma/(w\epsilon)$ then gives the typical error in the
interfacial region.  The surface tension corresponding to
Eq.~(\ref{eq:F}) is $\sigma=2\sqrt 2/3$.  We take $w=2\sqrt 2$ as a
typical measure for the interface width.

\begin{figure}[htb]
\centerline{ 
  \epsfxsize=2.4in 
  \epsfbox{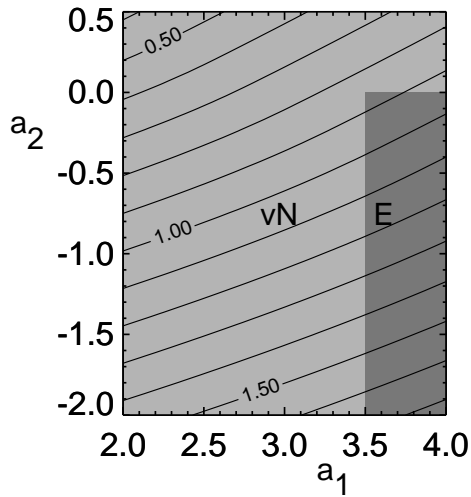} 
}
\caption{Contour of scaled error for a single direct update in a
$1024^2$ system. The systems are evolved well into the scaling regime
($t \approx 3000$) with a fixed-step Euler update. The errors are
found by comparing a single direct time step $\Delta t = A t^{1/3}$
with two steps of size $\dt/2$, and are then scaled by $2 \sigma/(A^2
w \epsilon)$ to estimate the average error magnitude per lattice site
in the interfacial region, as described in the text. }
\label{fig:contour}
\end{figure}

We have investigated this error for a variety of direct algorithms in
Fig.~\ref{fig:contour}, where we have plotted the interfacial error as
determined above divided by $A^2$.  We plot this error amplitude
against $a_1$ and $a_2$ for the same shaded regions [``vN'' and ``E'']
as identified in Fig.~\ref{fig:abstab}.  The typical interfacial error
for a given direct step of size $\dt = At^{1/3}$ may be obtained by
multiplying the appropriate contour value by $A^2$. 

\begin{figure}[htb]
\centerline{
\epsfxsize=3.0in
\epsfbox{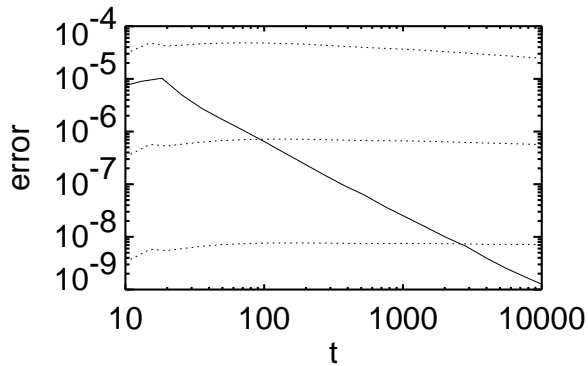}
}
\caption{Plot of scaled error per lattice site near the interface for
a single Euler step (solid), and for a single direct step with $a_1=3$
and $a_2=0$ (dotted with $\dt = A t^{1/3}$ where from bottom to top
$A=10^{-4}$, $10^{-3}$, and $10^{-2}$).  The scaling of the errors is
the same as in \protect\ref{fig:contour}, except that the errors are
not divided by $A^2$.  For the two smallest $A$ the scaling with $A^2$
is clearly seen, and so is the time independence of the error for the
driven direct step at later times. The system size is $2048^2$ and is
evolved with a Euler step with $\Delta t=0.05$.  }
\label{fig:driven}
\end{figure}

To illustrate the advantages of stable algorithms, as well as of a
detailed error analysis where it is possible, we show in
Fig.~\ref{fig:driven} how the error evolves in time for direct steps
with $\dt = At^{1/3}$ versus the Euler step with fixed $\dt$.  The
field $\phi$ is evolved by the Euler method, and during the evolution
error checking is done with single steps that do not contribute to the
evolution.  The decay of the Euler error shows that the Euler method
is asymptotically wastefully accurate.

\begin{figure}[htb]
\centerline{
\epsfxsize=3.25in
\epsfbox{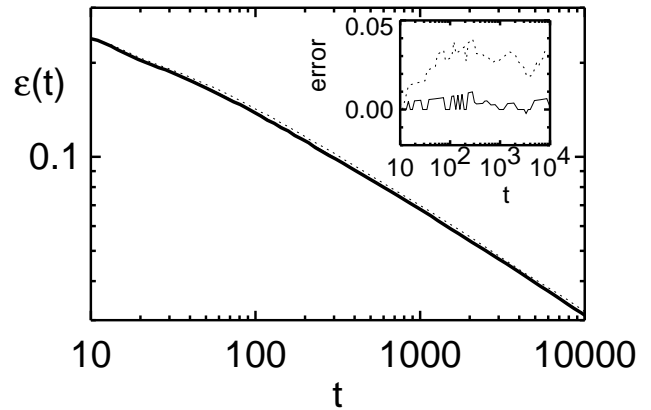}
}
\caption{Plot of $\epsilon$ versus $t$ for a Euler update (with $\dt =
0.05$, thick solid line) and with the evolution via a direct algorithm
($a_1=3$ and $a_2=0$) driven with $\dt = A t^{1/3}$ with $A
= 0.1$ (dotted line) and $0.01$ (thin solid line) in a $2048^2$ system.
Up until $t=10$ all systems were evolved with the Euler update.  In
the inset is plotted the percentage difference between the Euler and
direct updates: some error is introduced in the direct steps after
$t=10$ but at later times no increasing deviation from the Euler
evolution is seen.}
\label{fig:endrive}
\end{figure}

  Our single-step analysis and testing does not conclusively
demonstrate that an algorithm will be reasonably behaved under
successive steps, i.e.,\ there is a possibility of accumulation of
error.  In Fig.~\ref{fig:endrive} we show the free energy density for
systems evolved by a direct step and compare the evolution to that
obtained by the Euler method.  It appears that the errors do not
accumulate and the free energy decays properly as $t^{-1/3}$.

%%%%%%%%%%%%%%%%%%%%%%%%%%%%
\subsection{Toward $p>2$}
\label{subsec:pgtr2}

To go beyond the $p=2$ steps with $\dt\sim t^{1/3}$, it is necessary
to find a stable step that satisfies Eq.~(\ref{eq:CHp3}).  Comparing
with the stability conditions, Eqs.~(\ref{eq:vNstability}), we find
only marginally stable algorithms with $a_1=a_2=1/2$ and
$a_3=(3/2-m)/(3-m)$ for $0\leq m \leq 3$.  For $m=0$ this becomes the
Crank-Nicholson method, which as noted before, has a fixed time step
due to solvability considerations.  However, a marginal linear
iterative step is possible with $m=2$ and $a_3=-1/2$.  Unfortunately,
whether or not the marginality is a problem, the iterative method
(given by Eyre in \cite{Eyre98b}) fails to converge absolutely for
these parameters.  Evidently, then, it is not possible to construct a
useful $p=3$ step from the general step Eq.~(\ref{eq:CHgeneral}).

One possible way to develop a $p=3$ step is to use a method that is
both stable and second-order accurate in time.  For example, a
two-step method that uses both $\phi_{t-\dt}$ and $\phi_t$ to
determine the updated field $\phi_{t+\dt}$ can be made to have no
$O(\dt^2)$ error.  A preliminary study of vN stability for these
two-step methods indicates that these are a possibility.

It is worth considering the prospect of obtaining a $p\to\infty$ step:
according to the $p$ classification analysis this would allow the
natural $\dt\sim t^{2/3}$ time step.  However, the error terms need to 
be strictly proportional to $\partial_t^n\phi$ at each order $\dt^n$.
To achieve this with a one-step method one needs
\begin{equation}
   \tilde\phi_{t+\dt} - (1-a)\grad^2\mu_{t+\dt} = \phi_t + a\grad^2\mu_t.
   \label{eq:pinfty}
\end{equation}
Eq.~(\ref{eq:equala}) shows that this step will be linearly unstable
when $a > 1/2$ (for large enough $\dt$), while for $a<1/2$ one runs
into solvability problems.  At this point it seems unlikely that a
$p\to\infty$ algorithm for the CH equation will be possible.

%%%%%%%%%%%%%%%%%%%%%%%%%%%%%%%%%
\subsection{Asymptotic Scaling of Field Derivatives}
\label{subeq:fieldscaling}

In this section we derive the relations (\ref{eq:interfacedecay}) and
(\ref{eq:bulkdecay}) that provided the basis for $p$ classification.
Enough is known about CH dynamics that we can explicitly analyze the
leading asymptotic decay of mixed space and time derivatives to
arbitrary order.  We follow the review by Bray \cite{Bray94}, and we
restrict ourselves to the power-law scaling of these terms at
sufficiently late times, where all observable length scales that
describe the domain wall morphology, such as the interface curvature
radii, are proportional to the domain size $L\sim t^{1/3}$.  The
domain wall thickness $w$ does not grow with time, so $w \ll L$
asymptotically.  However, when analyzing the fields in the interfacial
region, defined as the locus of points within a distance $w$ of a
domain wall center (i.e., the surface $\phi=0$), both length scales
$L$ and $w$ can appear.  The remainder of the system is referred to
as the bulk.

The scale of the chemical potential $\mu$ is proportional to interface
curvature $\kappa$ due to the Gibbs-Thompson effect, and since $\kappa
\sim 1/L$
\begin{equation}
        \mu \sim 1/L \sim 1/t^{1/3}.
\end{equation} 
In the bulk, the chemical potential varies smoothly and continuously, 
so a Laplacian simply brings in more powers of $L$:
\begin{equation}
        \grad^2\mu \sim 1/L^3 \sim 1/t,
\end{equation} 
which implies $\partial_t\phi \sim 1/t$ via the equation of motion
(\ref{eq:CH2}).  Now we use the relation $\phi-\phi_{eq} \sim \mu$ in
the bulk \cite{Bray94} to relate derivatives of $\phi$ and $\mu$.  For
example, $\grad^2\phi\sim \grad^2\mu$, so $\partial_t\phi \sim
\grad^2\phi$.  Taking more time derivatives gives
\begin{equation} 
  \partial_t^n\phi \sim \grad^2 \partial_t^{n-1}\phi 
  \sim t^{-2/3}\partial_t^{n-1}\phi.
\end{equation}
Iterating this from the initial value for $\partial_t\phi$ gives
$\partial_t^n\phi \sim t^{-(1/3)-(2/3)n}$, the first term in
Eq.~(\ref{eq:bulkdecay}). 

When the time derivatives act on a power of the field $\phi^j$, the
resulting expression contains the $j$ fields and $n$ time derivatives
in various combinations.  In this case the asymptotic decay comes from
the single term proportional to $\phi^{j-1}\partial_t^n\phi$, which
means the decay for $\phi^j$ derivatives is the same as the $j=1$
case, since the field $\phi$ is order unity in the bulk.  To
illustrate, consider $\partial_t^2\phi^3 = 6\phi(\partial_t\phi)^2 +
3\phi^2\partial_t^2\phi$.  The second term decays as $t^{-5/3}$ as
advertised, while the first term goes as $(t^{-1})^2$ and is
asymptotically negligible.

Adding spatial derivatives in the bulk simply brings
more factors of $L^{-1}$, so
\begin{equation} 
  (\grad^2)^k\partial_t^{n-1}\phi^j 
  \sim L^{-2k}\partial_t^{n-1}\phi^j
  \sim t^{-2k/3} t^{-(1/3)-(2/3)n}
\end{equation}
which gives the second term in Eq.~(\ref{eq:bulkdecay}).

Near interfaces, $\phi$ changes by an amount $\Delta\phi_{eq}$ in the
amount of time, $\tau = w/v \sim t^{2/3}$, it takes an interface to
pass by.  Therefore we get $\partial_t\phi \sim t^{-2/3}$ in the
interfacial region, in contrast to $\partial_t\phi\sim t^{-1}$.  in
the bulk.  To determine the scaling $\partial_t^2\phi$, consider
sitting at a point just outside the interfacial region, in front of
the moving interface.  At a time $O(\tau)$ later this point
will be in the interfacial region, so $\partial_t\phi$ will have
changed from a bulk to an interfacial value.  This gives
\begin{equation}
  \partial_t^2\phi \sim (t^{-2/3} - t^{-1})/\tau \sim t^{-4/3}.
\end{equation}
Repeating this argument for higher derivatives gives $\partial_t^n\phi
\sim t^{-2n/3}$ in the interface, the first term in
Eq.~(\ref{eq:interfacedecay}).

For time derivatives of $\phi^j$ at the interface, we again get
multiple terms with the various combinations of $n$ time derivatives
and $j$ fields.  In this case, however, every term contributes to the
asymptotic decay.  Essentially every time derivative, wherever it
acts, brings a factor of $t^{-2/3}$, and these are the only factors
causing the decay.  Hence $\partial_t^n\phi^j \sim \partial_t^n\phi$.
Finally, adding spatial derivatives in the interfacial region brings
factors of $w^{-1}$ rather than $L^{-1}$, and so does not change the
asymptotic decay.  This proves the second relation in
Eq.~(\ref{eq:interfacedecay}).

%%%%%%%%%%%%%%%%%%%%%%%%%%%%%%%%%%%%%%%%%%%%%%%%%%%%%%%%%%%%%%%%%%%%%%%%%
\section{Computational Advantage}
\label{sec:speedup}

Having established the possibility of controlled accuracy CH
simulation with a growing step size $\dt\sim t^\alpha$, we now explore
the relative computational advantage offered by such an algorithm.  As
described in Sec.~\ref{sec:intro}, the goal in such simulations is to
evolve as far as close as possible to the scaling regime, meaning the
largest possible $L(t)$.  This means evolving until finite size
effects enter, since stopping earlier means a smaller system size
could be chosen.  Finite size effects are expected to appear when
$L(t)\sim L_0 t^{1/3}$ is some fraction of the system size, so we
define the simulation ending time $t_{max}$ by $L(t_{max}) = f
L_{sys}$, or
\begin{equation} 
  t_{max} = (f L_{sys}/L_0)^3 = (f \dx N/L_0)^3
  \label{eq:tmax}
\end{equation}
where $N$ is the linear size of the lattice and 
$f$ is a small constant factor.  There is some
arbitrariness in the definition of the length scale $L(t)$.  We take
the inverse interface density as our measure, that is
\begin{equation}
  L(t) = {L_{sys}^d\over A_{int}} = {\sigma\over\epsilon(t)} =
  {\sigma\over\epsilon_0}t^{1/3}
\end{equation}
using the interfacial area $A_{int}$ from
Sec.~\ref{subsec:directerror}, and its relation to the free energy
density and surface tension derived therein.  From our data in $d=2$
we find $\epsilon_0\simeq 0.675$, so we take $L_0 = \sigma/\epsilon_0
\simeq 1.40$.

Evolving to $t_{max}$ with the Euler step (or any fixed-size step)
requires $n= t_{max}/\dt_0$ steps, where $\dt_0$ is the step size.
For our square lattice with $\dx=1$ we find $\dt_0=0.05$ is close to
the maximum stable value.  More generally, one expects $\dt_0\sim
\dx^4$ \cite{Rogers88}.  Evolving to $t_{max}$ with a growing step
size $\dt\sim At^\alpha \sim dt/dn$ requires
\begin{equation}
    n = \int_{t_0}^{t_{max}} A^{-1} t^{-\alpha}dt 
    \sim {1\over A(1-\alpha)} t_{max}^{1-\alpha}
    \label{eq:stepnumber}
\end{equation}
where a fixed-size step is used until some time $t_0\ll t_{max}$, 
and we assume $t_0$-dependent terms are negligible.

Finally, we determine empirically the ratio of computer time per step
$\beta = \tau_{\rm stable}/\tau_{\rm Euler}$.  For direct steps, the
FFT involved implies $\beta\sim \log N$.  For lattices of size
$1024^2$ to $4096^2$ we find $\beta\simeq 2.3\pm 0.1$.

Putting all this together, we find the ratio of computer time cost
\begin{equation}
  {\hbox{Euler}\over\hbox{Stable}} = 
  {A(1-\alpha) t_{max}^\alpha\over\beta \dt_0} = 
  {A(1-\alpha)\over\beta\dt_0}\left(f\dx\over L_0\right)^{3\alpha} 
  N^{3\alpha}
\end{equation}
For direct steps, $\alpha=1/3$, so the relative speedup over Euler
integration grows with the system size as $N/\log N$.  From
$\dt_0\sim\dx^4$ we also see the speedup factor scaling as $1/\dx^3$,
making stable steps an optimal choice when a smaller lattice spacing
is desired.  A $p=3$ algorithm has $\alpha=4/9$ and offers a speedup
factor of $N^{4/3}/\log N$.

We conclude by plugging in reasonably conservative parameter values.
From Fig.~\ref{fig:contour} we see that the typical interfacial error
for the $a_1=3$, $a_2=0$ direct step is about $0.7A^2$.  This is to be
compared to $\Delta\phi_{eq}=2$, the range in which $\phi$ varies.
The choice $A=0.1$ is shown in Fig.~\ref{fig:endrive} to give an error
in the free energy density around 3\% of the Euler value.  While this
seems perhaps high, we note that this is comparable and probably
smaller than the error already introduced in the Euler discretization
of the continuum CH equation due to the large lattice constant.  It is
an interesting question for future study what choice of $\dx$ and $A$
will give optimal accuracy and efficiency.  We conclude that $A=0.1$
is a reasonable choice.  We also take $\alpha=1/3$, $f=1/10$,
$\beta=2.5$, $\dx=1$, and $L_0$ as given above.  These combine to give
a factor $0.038N$.  For a $1024^2$ lattice the direct step is a factor
40 faster than the Euler method, while for a $8192^2$ lattice it is a
factor 300 faster!

%%%%%%%%%%%%%%%%%%%%%%%%%%%%%%%%%%%%%%%%%%%%%%%%%%%%%%%%%%%%%%%%%%%%%%%%%
\section{Conclusions and Future Directions}
\label{sec:conclusion}

We have seen that the general Cahn Hilliard (CH) step,
Eq.~(\ref{eq:CHgeneral}), provides a range of linearly stable
algorithms that prove to be gradient stable for enormous single time
steps up to $\Delta t = 10^{10}$.  With these steps unphysical
instabilities arising from the discrete implementations are no longer
the limiting factor. Instead accuracy considerations dominate.  For
conserved Cahn Hilliard coarsening, we have analyzed and tested the
accuracy scaling for single dynamical time steps that {\em increase}
without bound with time as $\Delta t \sim t^\alpha$.  We find that the
errors are dominated at the order $\Delta t^p$ where they are no
longer proportional to $\partial_t^p\phi$.  These dominant errors
restrict the growth of the time step to grow as $\Delta t \sim
t^{2(p-1)/(3p)}$, which approaches the natural dynamical time step
$\tau \sim t^{2/3}$ only as $p \rightarrow \infty$.  The Euler method,
by contrast, is restricted to a constant $\Delta t$.  This is also the
case for existing implicit Fourier spectral algorithms.  The {\bf
direct} steps obtained from Eq.~(\ref{eq:CHgeneral}) with $a_3=1$ are
linear and diagonalized in Fourier space, and so can be simply
integrated via FFT's.  A range of parameters, described by the shaded
boxes in Fig.~\ref{fig:abstab}, are stable.  These direct steps
exhibit $p=2$ and so allow $\Delta t \sim t^{1/3}$, which results in
speedup factors proportional to the linear size of the system.

Future work in further developing these methods includes determining
possible $p=3$ algorithms, for which $\dt\sim t^{4/9}$ is possible
and the relative speedup over the Euler method is order $N^{4/3}/\log N$.
Our preliminary work has shown that $O(\dt^2)$ accurate two-step
methods can be made unconditionally vN stable.  It remains to test 
these stability predictions numerically to see if useful $p=3$ algorithms
are possible.  

It is straightforward to construct a Fourier spectral method integration
algorithm for the stable steps analyzed here.  In fact, the numerical
cost of the spectral method would be quite small, since
the direct steps already employ FFT's for solving the update
equation.  The primary benefit of the spectral method for unstable
algorithms is that it significantly enhances the maximum $\dt_0$
allowed by stability.  It is not clear how much benefit spectral methods
would bring to an already stable algorithm, but this should be explored.

With the Euler step, the simulation efficiency was strongly dependent
on $\dx$, leading to choosing values that were as large as feasible.
Consequently the interface profile is typically poorly resolved,
modifying and introducing significant anisotropy into the surface
tension.  In contrast, the efficiency of these stable methods is much
less dependent on the choice of lattice size, making them a useful
tool in applications where a more accurate interface profile is
desired.

Our analysis has been for errors after a single time-step. If the
single-step errors are small enough, the linear stability of bulk
solutions should control the errors from accumulating.  For the CH
equation at least, our observed $\epsilon \sim t^{-1/3}$ decay of the
free energy, even when $\dt \sim A t^{1/3}$, indicates that there is
no significant curvature-dependent modification of interfacial
speeds.  Nevertheless, it will be important to study the relationship
between single-step errors and errors of the asymptotic scaling
functions describing correlations to confirm this.

We feel that our basic approach should be applicable in a wide variety
of systems that have both nonlinearities and numerical instabilities.
There are just three basic ingredients: i) allow for a general
semi-implicit parametrization, following Eq.~(\ref{eq:CHgeneral}); ii)
check for unconditional von Neumann (linear) stability of an
individual update step, following Sec.~\ref{subsec:Eyrestability}; and
iii) numerically test the vN stable algorithms for speed, accuracy,
and nonlinear stability in order to pick the best parameters for
further study.  As long as the stability criteria are lattice
independent, the resulting algorithms should be applicable on any
regular lattice in any spatial dimension, and even on irregular
discretizations such as used in adaptive mesh techniques.

%%%%%%%%%%%%%%%%
\acknowledgments

BPV-L acknowledges financial support from {\it
Sonderforschungsbereich} 262 and the hospitality of the University of
Mainz, where part of this work was completed. ADR acknowledges
financial support through NSERC. We would like to thank Mowei Cheng,
David Eyre, Baruch Meerson, James Miante, Mathias Rauscher, and Jim
Sethna for stimulating discussions.

\appendix
%%%%%%%% START OF APPENDICES %%%%%%%%%%%%%%%%%%%%%%%%%%%%%%%%%%%%%%%%%%%%%%
\section{Eyre's Theorem}
\label{app:Eyre}

We repeat Eyre's stability theorem \cite{Eyre98a} here to flesh out
the derivation for the conserved dynamics case, and to clarify some
details of the proof. In particular, there are a few misleading
equations in \cite{Eyre98a} that lack factors of the norm of the
vector.  More substantively, we find that Eyre's theorem as originally
presented was slightly more restrictive than necessary.  Note that
questions of accuracy are not addressed in this proof, only questions
of numerical stability.

A central quantity in Eyre's theorem is the Hessian matrix
\begin{equation}
	M_{ij} = {\partial^2 F\over\partial\phi_i\partial\phi_j}
\end{equation}
where $F$ is the free energy and $\phi_i$ represents the field at the
lattice site $i$ (we consider only scalar one-component fields here).
For free energies of interest in coarsening, this matrix has both
positive and negative eigenvalues.  Eyre finds a stable first-order
step by splitting the free energy into {\it contractive} and {\it
expansive} parts, $F=F^C+F^E$, such that $F^C$ is convex and $F^E$ is
concave; that is, the eigenvalues of $M^C_{ij}$, the Hessian matrix
corresponding to $F^C$, are strictly non-negative, and the eigenvalues
of $M^E_{ij}$ corresponding to $F^E$ are strictly non-positive, for
any possible field configuration.

Let $\lambda_{min}<0$ represent the lower bound for the eigenvalues of
$M$ over all fields $\bphi$ (such a bound must exist \cite{Eyre98a}),
and $\lambda^E_{max}\leq 0$ represent the upper bound on the
eigenvalues of $M^E$.  The main result is that if
\begin{equation}
        \lambda^E_{max} \leq \frac 1 2\lambda_{min}
        \label{eq:evinequal}
\end{equation}
then the field equations of motion
\begin{equation}
        \phi_{t+\dt} + 
		\dt {\delta F^C\over\delta\phi}\bigg\vert_{\bphi_{t+\dt}} =
        \phi_t - \dt {\delta F^E\over\delta\phi}\bigg\vert_{\bphi_t}
        \label{eq:eyrestepNC}
\end{equation}
for nonconserved dynamics or
\begin{equation}
        \phi_{t+\dt}-
	\dt\grad^2{\delta F^C\over\delta\phi}\bigg\vert_{\bphi_{t+\dt}}=
        \phi_t + \dt\grad^2{\delta F^E\over\delta\phi}\bigg\vert_{\bphi_t}
        \label{eq:eyrestepC}
\end{equation}
for conserved dynamics lead to a strict non-increase of the free
energy in time:
\begin{equation}
        F(\bphi_{t+\dt}) \leq F(\bphi_t),
        \label{eq:eyreresult}
\end{equation}
where we have suppressed the lattice index for clarity.  This holds
unconditionally for all field configurations $\bphi_t$ and all step
sizes $\dt>0$.  Convexity of $F^C$ ensures that the implicit equation
for $\phi_{t+\dt}$ has a unique solution.

The energy dissipation property, along with other reasonable
requirements like positivity of $F$, is called {\it gradient
stability} by Eyre \cite{Eyre98a}. While gradient stability can be
obtained for many algorithms, such as the Euler step, by using a small
enough $\dt$, the algorithm defined by
Eqs.~(\ref{eq:evinequal})--(\ref{eq:eyrestepC}) guarantees it for {\em
arbitrarily large} $\dt$!  Even so, finding the splittings into $F^C$
and $F^E$ that lead to Eq.~(\ref{eq:evinequal}) can be a difficult
task, and the splittings, if they exist, may not be unique.

Condition Eq.~(\ref{eq:evinequal}) corrects the corresponding
condition in \cite{Eyre98a}, $\lambda^E_{max} \leq \lambda_{min}$.
The current form is less restrictive since $\lambda_{min}<0$.

An extremely useful corollary to Eyre's theorem is that if the
eigenvalue condition Eq.~(\ref{eq:evinequal}) is satisfied
for a restricted set of fields $\bphi$, then Eq.~(\ref{eq:eyreresult})
still applies for all $\dt$ provided $\bphi_t$ always stays within
this restricted set. For example, $\bphi$ could be field
configurations with $\phi_i^2 < \phi_0^2$ for all $i$, for some constant
$\phi_0$.  This can be useful when $\bphi$ is physically restricted by the
dynamics, and is employed in the {\bf direct} algorithms discussed in
Sec.~\ref{subsec:Eyrestability}

The proof of Eq.~(\ref{eq:eyreresult}) relies on two inequalities
\begin{equation}
        F(\bphi_{t+\dt}) - F(\bphi_t) \leq
        \sum_i \d\phi_i 
		{\partial F\over\partial\phi_i}\bigg\vert_{\bphi_{t+\dt}}
        - \frac 1 2 \lambda_{min} |\d\bphi|^2
        \label{eq:inequal1}
\end{equation}
and
\begin{equation}
        \sum_i \d\phi_i \biggl(
        {\partial F^E\over\partial \phi_i}\bigg\vert_{\bphi_{t+\dt}} -
        {\partial F^E\over\partial \phi_i}\bigg\vert_{\bphi_t}\biggr) \leq
        \lambda^E_{max}|\d\bphi|^2
        \label{eq:inequal2}
\end{equation}
where $\d\phi_i \equiv \phi_{i,t+\dt} - \phi_{i,t}$ and
$|\d\bphi|^2=\sum_i\d\phi_i^2$.  These are simply properties of
multivariable functions, and are derived in appendix
\ref{app:Calculus} for completeness.

Consider first {\it nonconserved dynamics}.  By adding $\dt[\partial
F^E/\partial\phi_i]_{\bphi_{t+\dt}}$ to both sides of the equation of
motion Eq.~(\ref{eq:eyrestepNC}) one obtains
\begin{equation}
        {\partial F\over\partial\phi_i}\bigg\vert_{\bphi_{t+\dt}}
        = -{1\over\dt}\d\phi_i +
        {\partial F^E\over\partial\phi_i}\bigg\vert_{\bphi_{t+\dt}} -
        {\partial F^E\over\partial\phi_i}\bigg\vert_{\bphi_t}.
        \label{eq:NCtheorem1}
\end{equation}
Substituting this into Eq.~(\ref{eq:inequal1}) gives
\begin{eqnarray}
        F(\bphi_{t+\dt}) - F(\bphi_t) &\leq&
         \sum_i \d\phi_i
        \biggl({\partial F^E\over\partial\phi_i}\bigg\vert_{\bphi_{t+\dt}}-
        {\partial F^E\over\partial\phi_i}\bigg\vert_{\bphi_t}\biggr)
         \nonumber\\
        && - \biggl(\frac 1 2 \lambda_{min} + {1\over\dt}\biggr) |\d\bphi|^2.
        \label{eq:NCtheorem2}
\end{eqnarray}
Next use Eq.~(\ref{eq:inequal2}) to complete the proof:
\begin{eqnarray}
        F(\bphi_{t+\dt})-F(\bphi_t) &\leq& \biggl(\lambda^E_{max}
        - \frac 1 2 \lambda_{min} - {1\over\dt}\biggr) |\d\bphi|^2, \nonumber\\
        &\leq& 0
\end{eqnarray}
where the last inequality follows by assumption
Eq.~(\ref{eq:evinequal}).

Analyzing {\it conserved dynamics} is complicated by the Laplacian in
the equations of motion.  Consider a general dimensional lattice of
$n$ sites with lattice Laplacian $(\grad^2)_{ij}\equiv A_{ij}$ a
symmetric $n\times n$ matrix with eigenvalues $\lambda_1=0$ and
$\lambda_m<0$ for all $m>1$.  Let $u^{(m)}_i$ represent the $i$th
component of the $m$th eigenvector of $A$, then we can write the
Kronecker delta function as
\begin{equation}
        \delta_{ik} = \sum_{m=1}^n u^{(m)}_i u^{(m)}_k =
        \sum_{j=1}^n \widetilde A_{ij} A_{jk}
        + u^{(1)}_i u^{(1)}_k
        \label{eq:Kronecker}
\end{equation}
where the pseudo-inverse $\widetilde A$ is defined by
\begin{equation}
        \widetilde A_{ij} = \sum_{m\neq 1}^n
        {1\over\lambda_m} u^{(m)}_i u^{(m)}_j.
\end{equation}
Note that the eigenvalue $\lambda_1=0$ corresponds to the eigenvector
$u^{(1)}_i=1/\sqrt n$ for all $i$, i.e., a uniform field.  Now we
insert Eq.~(\ref{eq:Kronecker}) into the sum in
Eq.~(\ref{eq:inequal1}) and sum on $k$ to get
\begin{eqnarray}
        F(\bphi_{t+\dt}) - F(\bphi_t) &\leq&
        \sum_{i,j,k} \d\phi_i \widetilde A_{ij} A_{jk}
        {\partial F\over\partial\phi_k}\bigg\vert_{\bphi_{t+\dt}}  \nonumber\\
        &-& \frac 1 2 \lambda_{min} |\d\bphi|^2
        \label{eq:Ctheorem1}
\end{eqnarray}
where we have used $\sum_i\d\phi_i=0$, which follows from the
conservation law.  Proceeding by analogy with the nonconserved case,
we subtract $\dt A_{jk}[\partial F^E/\partial\phi_k]_{\bphi_{t+\dt}}$
from both sides of the equation of motion Eq.~(\ref{eq:eyrestepC}) to
get
\begin{eqnarray}
        &&\sum_k A_{jk}{\partial F\over\partial\phi_k}
        \bigg\vert_{\bphi_{t+\dt}} = \nonumber\\ 
	&& {\delta\phi_j\over\dt}  + \sum_k A_{jk}\biggl(
        {\partial F^E\over\partial\phi_k}\bigg\vert_{\bphi_{t+\dt}}-
        {\partial F^E\over\partial\phi_k}\bigg\vert_{\bphi_t}\biggr)
\end{eqnarray}
Substituting this into Eq.~(\ref{eq:Ctheorem1}) gives
\begin{eqnarray}
        F(\bphi_{t+\dt}) &-& F(\bphi_t) \leq
         \sum_i \d\phi_i
        \biggl({\partial F^E\over\partial\phi_i}\bigg\vert_{\bphi_{t+\dt}}-
        {\partial F^E\over\partial\phi_i}\bigg\vert_{\bphi_t}\biggr)
         \nonumber\\
         &-& \frac 1 2 \lambda_{min} |\d\bphi|^2 + 
	 {1\over\dt}\sum_{i,j}\d\phi_i\d\phi_j \widetilde A_{ij}
        \label{eq:Ctheorem2}
\end{eqnarray}
which is identical to Eq.~(\ref{eq:NCtheorem2}) except for the $1/\dt$
term.  From the definition of $\widetilde A$ and an expansion of
$\d\bphi$ in the eigenvalues $\b u^{(m)}$ it follows that
\begin{equation}
        \sum_{i,j}\d\phi_i \d\phi_j \widetilde A_{ij} \leq 0
\end{equation}
so this term can be dropped from the right hand side of
Eq.~(\ref{eq:Ctheorem2}) and the proof follows as before to yield
Eq.~(\ref{eq:eyreresult}).

%%%%%%%%%%%%%%%%%%%%%%%%%%%%%%%%%%%%%%%%%%%%%%%%%%%%%%%%%%%%%%%%%%%%%%%%%%
\section{Inequalities used in Eyre's Theorem}
\label{app:Calculus}

For completeness, we re-derive Eqs.~(\ref{eq:inequal1}) and
(\ref{eq:inequal2}) here.  Consider a general function $f(\b x)$ of
$n$ variables $\b x=(x_1,\dots,x_n)$.  From the Fundamental Theorem of
Calculus
\begin{equation}
        f(\b x+\b y) - f(\b x) = \sum_i y_i \int_0^1 ds_1
        {\partial f\over\partial x_i}\biggr\vert_{\b x+s_1\b y},
\end{equation}
that is, we introduce the parameter $s_1$ to integrate along the
``diagonal'' path from $\b x$ to $\b x+\b y$.  Similarly, we can write
\begin{equation}
  {\partial f\over\partial x_i}\bigg\vert_{\b x+s_1\b y} -
  {\partial f\over\partial x_i}\bigg\vert_{\b x} =
  \sum_j y_j\int_0^{s_1} ds_2
  {\partial^2 f\over\partial x_i\partial x_j}\bigg\vert_{\b x+s_2\b y}.
  \label{eq:firstderiv}
\end{equation}
Combining these gives the identity
\begin{eqnarray}
        &&f(\b x+\b y) - f(\b x) =
        \sum_i y_i {\partial f\over\partial x_i}\bigg\vert_{\b x} \nonumber\\
        &&\qquad+ \int_0^1 ds_1\int_0^{s_1} ds_2 \sum_{i,j} y_i y_j
        {\partial^2 f\over\partial x_i\partial x_j}\bigg\vert_{\b x+s_2\b y}
        \label{eq:fidentity}
\end{eqnarray}
Now consider the case where the eigenvalues of the matrix
$M_{ij}=\partial^2 f/\partial x_i\partial x_j$ are bounded from below
by some constant $\lambda_{min}$ for all $\b x$.  In this case
\begin{equation}
        \sum_{i,j} y_i y_j
        {\partial^2 f\over\partial x_i\partial x_j}\bigg\vert_{\b x+s_2\b y}
        \geq \lambda_{min}|\b y|^2
        \label{eq:Minequal}
\end{equation}
which follows straightforwardly from an expansion of $\b y$ in the
basis of eigenvectors of $M$, with $|\b y|^2=\sum_i y_i^2$.  Thus we
have
\begin{equation}
        f(\b x+\b y) - f(\b x) \geq
        \sum_i y_i {\partial f\over\partial x_i}\bigg\vert_{\b x}
        + \frac 1 2 \lambda_{min}|\b y|^2
        \label{eq:inequal}
\end{equation}
where the $1/2$ follows from the $s$ integrals.  Taking the function
$f$ to be the free energy $F$ with $\b x=\bphi_{t+\dt}$ and $\b
y=\bphi_t-\bphi_{t+\dt}$ results in Eq.~(\ref{eq:inequal1}).

The second inequality results from setting $s_1=1$ in
Eq.~(\ref{eq:firstderiv}), then multiplying by $y_i$ and summing
\begin{equation}
        \sum_i y_i\biggl({\partial f\over\partial x_i}\bigg\vert_{\b x+\b y} -
        {\partial f\over\partial x_i}\bigg\vert_{\b x}\biggr) =
        \sum_{i,j} y_i y_j\int_0^1 ds
       {\partial^2 f\over\partial x_i\partial x_j}\bigg\vert_{\b x+s\b y}.
\end{equation}
We then use a relation similar to Eq.~(\ref{eq:Minequal}), only with
the eigenvalues of $\partial^2f/\partial x_i\partial x_j$ assumed to
be bounded {\it above} by $\lambda_{max}$, to get
\begin{equation}
        \sum_i y_i\biggl({\partial f\over\partial x_i}\bigg\vert_{\b x+\b y} -
        {\partial f\over\partial x_i}\bigg\vert_{\b x} \biggr)
        \leq \lambda_{max}|\b y|^2.
        \label{eq:inequalAP2}
\end{equation}
Now we can take $f=F^E$ and $\b x$ and $\b y$ as before to get 
Eq.~(\ref{eq:inequal2}).

%%%%%%%%%%%%%%%%%%%%%%%%%%%%%%%%%%%%%%%%%%%%%%%%%%%%%%%%%%%%%%%%%%%%%%%%%%


\begin{thebibliography}{}
\bibitem{Bray94}   A.~J.~Bray, Adv. Phys. {\bf 43}, 357 (1994).  
%
\bibitem{Rogers88} T.~M.~Rogers, K.~R.~Elder, and R.~C.~Desai,
Phys. Rev. B {\bf 37}, 9638 (1988).
%
\bibitem{Eyre98a} D.~J.~Eyre, ``An Unconditionally Stable One-Step
Scheme for Gradient Systems,''
preprint. (\url{http://www.math.utah.edu/~eyre/research/methods/stable.ps})
%
\bibitem{Eyre98b} D.~J.~Eyre, in {\em Computational and Mathematical
Models of Microstructural Evolution}, edited by J.~W.~Bullard {\em et
al.} (The Materials Research Society, Warrendale, PA, 1998),
pp. 39--46.
%
\bibitem{Provatas98} N.~Provatas, N.~Goldenfeld, and J.~Dantzig,
Phys. Rev. Lett.  {\bf 80}, 3308 (1998).
%
\bibitem{Oono88} Y.~Oono and S.~Puri, Phys. Rev. A {\bf 38}, 434
(1988).
%
\bibitem{Chen98} L.-Q.~Chen and J.~Shen, Comput.\ Phys.\ Commun.\ {\bf
108}, 147 (1998).
%
\bibitem{Zhu99} J.~Zhu, L.-Q.~Chen, J.~Shen, and V.~Tikare,
Phys. Rev. E {\bf 60}, 3564 (1999).
%
\bibitem{subsaturation} For regions not near an interface, the field
is given by $\phi - \phi_{eq} \propto \mu$.  For the interior regions
of curved interfaces, $\mu$ and $\phi_{eq}$ have the same sign, and
thus result in a value $|\phi| > |\phi_{eq}|$.  Note that this excess
is of order $t^{-1/3}$.
%
\bibitem{NumericalRecipes} W.~H.~Press, S.~A.~Teukolsky,
W.~T.~Vetterling, and B.~P.~Flannery, {\it Numerical Recipes in C},
2nd ed. (Cambridge University Press, New York, 1993).
\end{thebibliography}
\end{document}